\documentclass[amssymb,amsmath,superscriptaddress,footinbib,twocolumn]{revtex4}
\usepackage{natbib,graphicx,subeqnarray,fancyhdr,amsmath,multirow,color, mathrsfs,amsfonts}

\providecommand\bnabla{\boldsymbol{\nabla}}
\providecommand\bcdot{\boldsymbol{\cdot}}

\providecommand\bu{\boldsymbol{u}}

\providecommand\Pen{\mbox{\textit{Pe}}}
\providecommand\Da{\mbox{\textit{Da}}}

\newcommand\dd{\mathrm{d}}
\newcommand\ee{\mathrm{e}}
\providecommand\unit{\boldsymbol{\hat{\imath}}}

\newcommand{\pd}[2]{\frac{\partial #1}{\partial #2}}

\begin{document} 
\title{Phoretic self-propulsion at large P\'eclet numbers}
\author{ Ehud Yariv}
\email{udi@technion.ac.il}
\affiliation{Department of Mathematics, Technion --- Israel Institute of Technology, Haifa 32000, Israel}
\author {S\'ebastien Michelin}
\email{sebastien.michelin@ladhyx.polytechnique.fr}
\affiliation{LadHyX, D\'epartement de M\'ecanique, Ecole Polytechnique --- CNRS, 91128 Palaiseau, France}

\date{\today}


\begin{abstract}
We analyse the self-diffusiophoresis of a spherical particle animated by a nonuniform chemical reaction at its boundary. We consider two models of solute absorption, one with a specified distribution of interfacial solute flux, and one where this flux is governed by first-order kinetics with a specified distribution of rate constant. We employ a macroscale model 
where the short-range  interaction of the solute with the particle boundary is represented by an effective slip condition. The solute transport is governed by an advection--diffusion equation.

We focus upon the singular limit of large P\'eclet numbers, $\Pen\gg1$.
In the fixed-flux model, the excess-solute concentration is confined to a narrow boundary layer. The scaling pertinent to that limit allows to decouple the problem governing the solute concentration from the flow field. The resulting nonlinear boundary-layer problem is handled using a transformation to stream-function coordinates and a subsequent application of Fourier transforms, and is thereby reduced  to a nonlinear
 integral equation governing the interfacial concentration. Its solution provides the requisite approximation for the particle velocity, which scales as $\Pen^{-1/3}$.
 
In the fixed-rate model, large P\'eclet numbers may be realized in different limit processes. We consider the case of large swimmers or strong reaction, where the Damk\"ohler number $\Da$ is large as well, scaling as $\Pen$. In that double limit, where no boundary layer is formed, 
we obtain a closed-form approximation for the particle velocity, expressed as a nonlinear functional of the rate-constant distribution; this velocity scales as $\Pen^{-2}$. Both the fixed-flux and fixed-rate asymptotic predictions agree with the numerical values provided by computational solutions of the nonlinear transport problem. 
\end{abstract}
\maketitle

\section{Introduction}
When a nonuniform chemical reaction takes place on the boundary  of a colloidal particle, the interaction of the reactants with that boundary results in a flow field and a consequent motion of the freely suspended particle. 
When the solute is electrically neutral, its short-range interaction with the particle boundary may be represented by  an effective slip condition, relating the fluid velocity at the outer edge of the interaction layer to the tangential gradient of solute concentration \citep*{Anderson:82}. A simple ``continuum'' model of such a slip-based self-diffusiophoresis was provided by Golestanian \emph{et al.} \cite*{Golestanian:07}, who for simplicity described the chemical reaction by a prescribed distribution of solute flux. A different ``colloidal'' description for this problem was also proposed by Ref.~\cite{Cordova:08}; its linkage to the continuum approach appears to be controversial \citep{Julicher:09,Brady:11}. 

A key assumption in the model of  Ref.~\cite{Golestanian:07} is the neglect of solute advection, resulting in a linear transport problem governing the solute concentration. 
It appears that the first systematic analysis of solute advection in the continuum description was carried out by Michelin \& Lauga \cite{Michelin:14}, who considered the relatively simple configuration of a spherical particle with an axially symmetric distribution of chemical reactions. They  considered two separate models of solute production: in the first,
following Ref.~\cite{Golestanian:07}, the flux of solute is prescribed along the particle boundary; in the second, following Ref.~\cite{Brady:11}, the rate constant associated with a first-order chemical reaction is specified there. 
Starting from the exact microscale model together with either a `fixed-flux' or `fixed-rate' boundary conditions, 
and using coarse-graining techniques familiar from electrokinetic analyses \citep{Yariv:09,Schnitzer:12:Dukhin},
Ref.~\cite{Michelin:14} derived a  macroscale model where the solute--boundary interaction is systematically transformed into an effective slip condition (cf. Ref.~\cite[][]{Anderson:82}). 

Michelin \& Lauga \cite{Michelin:14} solved their macroscale problem for finite values of the P\'eclet number $\Pen$  using computations. 
For the fixed-flux variant of that problem, these computations indicate that at large $\Pen$ the particle speed scales as $\Pen^{-1/3}$. While that scaling was already predicted by \cite{Julicher:09:generic}, the associated boundary-layer problem has not been solved. In this paper we address this singular problem using a boundary-layer analysis. This is supplemented by a  large-P\'eclet-number analysis of the comparable fixed-rate problem. 
In both problems we derive asymptotic approximations for the particle velocity, and compare them with the 
 values obtained from the computational solution of the respective macroscale models.
 
\section{Problem formulation}
We consider first the fixed-flux variant of the general problem described by Ref.~\cite{Michelin:14}. 
A sphere of radius $a$ is suspended in a solution of an otherwise uniform concentration, say $\mathscr{C}_\infty$. Due to 
surface reaction, the particle emits solute at a prescribed rate $\mathscr{A}$, which in general varies along the particle boundary; it is assumed axially symmetric. The interaction energy between the solute and the boundary, characterized by the lengthscale $\lambda$, is comparable with the thermal energy $kT$. Our goal is the calculation of the steady-state velocity acquired by the particle due to a nonuniform boundary distribution of $\mathscr{A}$. 
The characteristic value of the excess-solute concentration, relative to the ambient value, is $\mathscr{C}=\| \mathscr{A} \| a/D$, where $\| \mathscr{A} \|$ is some representative value of $\mathscr{A}$ and $D$ is the solute diffusivity. 
The interaction of the solute with the particle boundary is represented as a body force on the fluid, whose magnitude is of order $kT\mathscr{C} / a$; balancing this force with the viscous stress in the $O(\lambda)$-wide interaction layer provides the pertinent velocity scale, namely  
\begin{equation}
\frac{\| \mathscr{A} \| kT \lambda^2}{\eta D}, \label{u scale}
\end{equation}
$\eta$ being the solution viscosity.

We employ the macroscale formulation of Ref.~\cite{Michelin:14},  
appropriate to the limit where the interaction thickness $\lambda$ is small compared with $a$. In that formulation the pertinent fields are the flow and excess-solute concentration outside the interaction layer. 
In the dimensionless notation of Ref.~\cite{Michelin:14}  distances are normalized by $a$, the excess-solute concentration by $\mathscr{C}$, and the velocity scale $\mathscr{U}$ is chosen comparable to \eqref{u scale} (see below); the governing equations are written in a reference frame moving with the particle, where the problem is steady, using spherical coordinates in which the radial distance $r$ is measured from the sphere centre and the zenith angle $\theta$ is measured from the symmetry axis: see figure~\ref{fig1}(a). For convenience we actually use the alternative coordinate $\mu=\cos\theta$. 

The excess concentration $c$ is governed by the advection--diffusion equation,
\begin{equation}
\Pen \,\bu \bcdot \bnabla c = \nabla^2 c, \label{advection--diffusion}
\end{equation}
where 
\begin{equation}
\Pen = \frac{a \mathscr{U}}{ D} \label{Pe}
\end{equation}
is the P\'eclet number and $\bu$ the velocity field in the fluid; 
the imposed-flux condition,
\begin{equation}
\pd{c}{r} = k(\mu) \quad \text{at}\quad r=1, \label{flux}
\end{equation}
where $k = -\mathscr{A} / \| \mathscr{A} \|$ is an $O(1)$ scaled activity; and the attenuation condition,
\begin{equation}
c\to 0 \quad \text{as} \quad r \to \infty. \label{far c}
\end{equation}

The velocity field $\bu$ is governed by the Stokes equations; the far-field approach to a uniform velocity of magnitude $\mathcal{U}$ (see figure~\ref{fig1}(a)),
\begin{equation}
\bu \to - \unit \, \mathcal{U}  \quad \text{as} \quad r \to \infty,
\end{equation}
wherein $\unit$ is a unit vector in the $\theta=0$ direction and $\mathcal{U}$ is the velocity of the particle relative to the fluid; and the condition that the particle is force-free. 
The flow is animated by the slip condition, 
\begin{equation}
\bu = M  \bnabla_s c\quad \text{at}\quad r=1, \label{slip}
\end{equation}
wherein $M$ is the diffusio-osmotic slip coefficient and $\bnabla_s$ the surface-gradient operator. For an isotropic interaction potential $M$ is uniform, but may be either positive or negative; the scale $\mathscr{U}$ was chosen by Ref.~\cite{Michelin:14} to make $M$ of unity magnitude, hence
\begin{equation}
M=\pm1.
\end{equation}

The goal is the calculation of the swimming velocity $\mathcal{U}$ as a function of $\Pen$ and the distribution $k(\mu)$. Because of the standard structure of a slip-driven Stokes-flow problem, $\mathcal{U}$ is obtained using the reciprocal theorem  \citep{Brenner:64,Stone:96}, 
\begin{equation}
\mathcal{U} = -\frac{1}{4\pi} \unit \bcdot \oint_{r=1}  \bu \,\dd A \label{recip}
\end{equation}
where 
$\dd A$ is a dimensionless areal element (normalized by $a^2$). Use of the axisymmetric slip \eqref{slip} followed by integration by parts thus provides the formula 
\begin{equation}
\mathcal{U} = -M  \int_{-1}^1  \mu \, c(r=1,\mu)  \,\dd\mu. \label{recip in c}
\end{equation}
While the particle speed depends only upon the surface concentration of $c$, one cannot avoid in general the
solution of the flow problem: the concentration field is affected by the flow through the advection term in \eqref{advection--diffusion}. Since this coupling between the flow and solute concentration is nonlinear, so is the dependence of $\mathcal{U}$ upon both 
$\Pen$ and $k(\mu)$. 

Note that, in contrast to classical forced-convection problems \citep{Acrivos:65}, the P\'eclet number here does not involve an externally imposed velocity scale. Rather, it represents the intensity of interfacial chemical activity. Specifically, substitution of \eqref{u scale} into \eqref{Pe} implies that $\Pen$ is of order
\begin{equation}
\frac{\| \mathscr{A} \| kT \lambda^2 a}{\eta D^2}. \label{Pe in A}
\end{equation}
Michelin \& Lauga \cite{Michelin:14} used computations to analyse self-propulsion at 
arbitrary P\'eclet numbers and regular perturbations to obtain analytic approximation for small P\'eclet numbers.
In what follows, we employ singular perturbations to address the opposite asymptotic limit, $\Pen\gg1$.
\section{Large-P\'eclet-number limit}
The advection--diffusion equation \eqref{advection--diffusion} suggests that in the large-P\'eclet-number limit the solute concentration is dominated by advection, so at leading order 
\begin{equation}
\bu \bcdot \bnabla c = 0. \label{advection dominated}
\end{equation}
Since the flow is steady, this implies that $c$ is a constant along the streamlines of the leading-order flow. Anticipating  open streamlines, which originate and end at infinity,
condition \eqref{far c} then implies
\begin{equation}
c \equiv 0. \label{c is 0}
\end{equation}

The nil result \eqref{c is 0} is clearly incompatible with \eqref{flux}. This non-uniformity is associated with the singular nature of the large-P\'eclet-number limit, where the highest derivative is effectively multiplied by a small parameter. It suggests that a diffusive boundary layer is formed about the boundary $r=1$; in that thin layer, advection and diffusion of solute are comparable. Note that the Stokes flow is coupled to the solute concentration $c$ only through its value at $r=1$,
and is accordingly `unaware' of the steep gradient of $c$ near that surface; consequently, no boundary layer
exists in the flow field.  
\subsection{Boundary-layer formulation}
\label{subsec:Boundary-layer formulation}
We denote the thickness of the layer by $\delta (\ll1)$. Its scaling with $\Pen$ is readily obtained using dominant balances  \citep{Julicher:09:generic,Michelin:14}: from condition \eqref{flux} we see that $c=O(\delta)$ in the layer. The slip condition \eqref{slip} then implies the same scaling for the tangential velocity-component $v$, while the continuity equation in conjunction with the impermeability condition implies that the radial velocity-component $u$ is of order $\delta^2$. The advection--diffusion equation \eqref{advection--diffusion} thus yields $\delta = \Pen^{-1/3}$. It follows that in the boundary layer $c$ is $O(\Pen^{-1/3})$; the velocity field $\bu$ is of the same magnitude in the entire fluid domain, and then, given \eqref{recip}, so is also $\mathcal{U}$.

The above scaling suggests the expansions $u = \Pen^{-1/3} U + \cdots$ and $v = \Pen^{-1/3} V + \cdots$ for the velocity components,
where the rescaled components are $O(1)$ functions of $r$ and $\mu$. Within the boundary layer we use the expansion
\begin{equation}
c = \Pen^{-1/3} C + \cdots, \label{def C}
\end{equation}
where the rescaled concentration $C$ is an $O(1)$ function of the stretched radial coordinate
\begin{equation}
Y = \Pen^{1/3} (r-1) \label{Y}
\end{equation}
and $\mu$. In terms of the rescaled components, condition \eqref{slip} becomes
\refstepcounter{equation}
$$
U=0, \quad V = -M (1-\mu^2)^{1/2} f(\mu)  \qquad \text{at}\quad r=1,
\label{slip for U V}
\eqno{(\theequation{\mathit{a},\mathit{b}})}
$$
where
\begin{equation}
f(\mu) \stackrel{\mathrm{def}}{=} \left.\pd{C}{\mu}\right|_{Y=0}. \label{f}
\end{equation}

Since $U$ and $V$ vary on an $O(1)$ scale they are approximated in the boundary layer using Taylor expansions about $r=1$, which, in view of (\ref{slip for U V}\textit{a}), read
\begin{equation}
U = (r-1) \left.\pd{U}{r}\right|_{r=1} + \cdots, \quad V = \left.V\right|_{r=1} + \cdots.
\end{equation}
Use of the leading-order continuity equation, 
\begin{equation}
\pd{U}{r} = \pd{}{\mu}[(1-\mu^2)^{1/2}V],
\end{equation}
allows to express both approximations in terms of $\left.V\right|_{r=1}$.  
Substitution into the advection--diffusion equation \eqref{advection--diffusion} and making use of (\ref{slip for U V}\textit{b}) thus yields the parabolic equation,
\begin{equation}
M\pd{^2C}{Y^2} = (1-\mu^2) f(\mu) \pd{C}{\mu} - Y \pd{}{\mu}[(1-\mu^2) f(\mu)]\pd{C}{Y},
\label{advection--diffusion C}
\end{equation}
governing the boundary-layer solute concentration. It is supplemented by the boundary condition (cf.~\eqref{flux}) 
\begin{equation}
\pd{C}{Y} = k(\mu) \quad \text{at} \quad Y=0 \label{flux C}
\end{equation}
and the matching condition (see~\eqref{c is 0})
\begin{equation}
C \to 0 \quad \text{as} \quad Y\to\infty. \label{C decay}
\end{equation}
The boundary-value problem \eqref{advection--diffusion C}--\eqref{C decay} is uncoupled to that governing the flow, and may be solved independently. In view of \eqref{recip in c} there is actually no need to solve for the flow: once $C$ is known, the particle velocity is obtained from \eqref{recip in c} and \eqref{def C}.
Note  however that with  $f(\mu)$ being related to $C$ through \eqref{f}, equation \eqref{advection--diffusion C} is a nonlinear one. 

The preceding boundary-layer problem appears similar to that  governing the nutrient concentration about a squirming sphere  \citep*{Magar:03,Michelin:11}. In the feeding problem, however, the slip distribution is prescribed, and is accordingly independent of the P\'eclet number; this results in a linear problem with a boundary-layer thickness that scales as the $-1/2$ power of that number.
\subsection{Boundary-layer analysis}
 Following the analysis of a similar boundary-layer problem \citep*{Schnitzer:14:bubble}, our approach in confronting \eqref{advection--diffusion C}--\eqref{C decay} is to calculate $C$ as if 
$f(\mu)$ were a prescribed quantity. The solution of the linear system \eqref{advection--diffusion C}--\eqref{C decay} in conjunction with \eqref{f} then provides a nonlinear equation governing $C$ at $Y=0$. Given \eqref{recip in c} and \eqref{def C}, this interfacial distribution is all that we need to obtain the swimming velocity.  

Our scheme is appropriate for the class of problems where the boundary layer does not detach; given \eqref{slip} this necessitates that on the boundary $C$ is strictly increasing or decreasing function of $\mu$; with no loss of generality we assume the latter,
\begin{equation}
f(\mu)<0. \label{f<0}
\end{equation}
Given \eqref{flux C} and the vanishing of $C$ at the outer edge of the boundary layer, it appears plausible that for
\eqref{f<0} to be satisfied $k(\mu)$ must be strictly increasing.

The parabolic equation \eqref{advection--diffusion C} does not appear amenable to integral transform. We accordingly shift to stream-function coordinates \citep{Levich:book}, replacing the independent variables $(Y,\eta)$ by $(\varPsi,\eta)$, with
\begin{equation}
\varPsi = -Y(1-\mu^2) f(\mu)
\end{equation}
(which is proportional to the Stokes stream-function near $r=1$.) 
Substituting this change of variables into \eqref{advection--diffusion C} we find that $C(\varPsi,\mu)$ satisfies the 
equation 
\begin{equation}
\pd{C}{\mu} = M (1-\mu^2) f(\mu) \pd{^2C}{\varPsi^2}, \label{advection--diffusion C Psi}
\end{equation}
while condition \eqref{flux C} becomes
\begin{equation}
\pd{C}{\varPsi} = -\frac{k(\mu)}{(1-\mu^2) f(\mu)} \quad \text{at} \quad \varPsi=0; \label{flux Psi}
\end{equation}
given \eqref{f<0}, the decay condition \eqref{C decay} now applies as $\varPsi\to\infty$:
\begin{equation}
C \to 0 \quad \text{as} \quad \varPsi\to\infty. \label{C decay Psi}
\end{equation}

The  boundary-value problem \eqref{advection--diffusion C Psi}--\eqref{C decay Psi}  is naturally solved using a Fourier cosine transform, defined as
\begin{equation}
\hat{C}(\omega,\mu) = \left(\frac{2}{\pi}\right)^{1/2} \int_0^\infty  C(\varPsi,\mu) \cos( \omega\varPsi) \,\dd \omega .
\end{equation}
Application of this transform to \eqref{advection--diffusion C Psi}--\eqref{C decay Psi} results in the first-order
equation
\begin{equation}
\pd{\hat{C}}{\mu} +M \omega^2  (1-\mu^2) f(\mu)\hat{C} = M \left(\frac{2}{\pi}\right)^{1/2} k(\mu). \label{ode}
\end{equation}
The complementary  solution of \eqref{ode}  is $A(\omega)\ee^{-M \omega^2 \chi(\mu)}$,
where
\begin{equation}
\chi(\mu) = \int_0^\mu f(q) (1-q^2) \, \dd q ; \label{xi}
\end{equation}
given \eqref{f<0}, $\chi(\mu)$ is strictly decreasing. A particular integral of \eqref{ode} is 
\begin{equation}
M  \left(\frac{2}{\pi}\right)^{1/2}  \int_{\mu_0}^\mu  k(\xi) \ee^{-M \omega^2 [\chi(\mu)-\chi(\xi)]} \, \dd\xi.
\label{particular}
\end{equation} 

The value of $\mu_0$ (and then $A$) is determined by the condition that $\hat{C}\to0$ as $\omega\to\infty$ regardless of the value of $\mu$. We now show that, upon choosing $\mu_0=M$, this condition yields $A=0$. Indeed, since the function $\chi$ is strictly decreasing this choice results in a negative power in the exponent appearing in \eqref{particular}, whereby the {particular} solution vanishes as $\omega\to\infty$. Given \eqref{f<0}, 
the sign of $\chi(\mu)$ is opposite to that of $\mu$, implying that the exponent in the complementary solution diverges
as $\omega\to\infty$ for a finite interval of $\mu$-values ($0<\mu<1$ for $M=1$; $-1<\mu<0$ for $M=-1$). Since $\hat{C}$ must vanish in that limit for all $\mu$, it follows that $A=0$. 

We conclude that $\hat{C}$ is provided by \eqref{particular} with $\mu_0=M$.  Applying the inverse transform, interchanging the order of integrations, and evaluation at $\varPsi=0$ we obtain, upon 
substitution of \eqref{f} and \eqref{xi}, the following equation governing $C$ on the particle boundary (denoted $C(\mu)$ for brevity):
\begin{equation} 
C(\mu) =  \frac{M}{\sqrt{\pi}} \int_{M}^\mu \frac{ k(\xi)\,\dd \xi }{\displaystyle \sqrt{M\int_\xi^\mu (1-q^2) \frac{\dd C}{\dd q} \,\dd q }}.
\label{C final}
\end{equation}
Since $C(\mu)$ is strictly decreasing, the local slip is directed in the $\pm\theta$ direction for $M=\pm1$. Given \eqref{recip in c}, this implies that the particle moves in the $\pm \unit$ direction. In the particle-fixed reference frame, the direction of the incident flow is then $\mp \unit$. In view of the parabolic nature of the boundary-layer problem, the value of $C$ at a certain value of $\mu$ is only affected by the distribution of $C$ upstream. 
For $M=1$ this upstream distribution corresponds to the interval $(\mu,1)$, while for $M=-1$ it corresponds to 
$(-1,\mu)$. Equation \eqref{C final} is indeed consistent with this information-propagation property.
\begin{figure*}
\begin{center}

\includegraphics[width=0.85\textwidth]{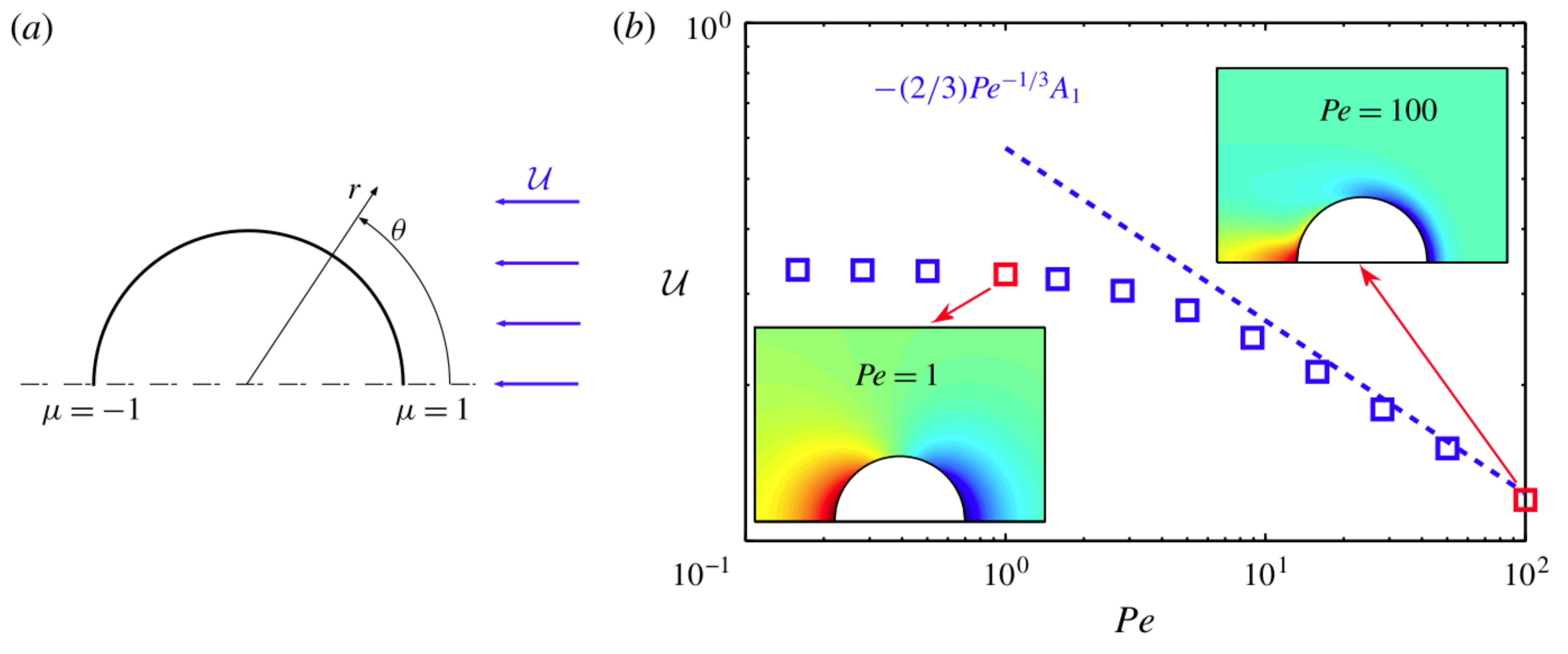}
\caption{(a) Schematic. (b) Particle velocity $\mathcal{U}$ for the fixed-flux model, with $k(\mu)=\mu$ and $M=1$. Dashed line: large-$\Pen$ approximation \eqref{final U}; symbols: computational results. The insets depict the solute-concentration distribution (scaled by the maximum surface concentration) as provided by the numerical solutions for $\Pen=1$ and $\Pen=100$.}
\label{fig1}
\end{center}
\end{figure*}

Equation \eqref{C final} constitutes an integral equation governing $C(\mu)$ \citep[cf.][]{Acrivos:57}. Upon expanding $C(\mu)$ in Legendre polynomials, $C(\mu)= \sum_{n=0}^\infty A_n P_n(\mu)$, it is transformed into a nonlinear algebraic system governing the coefficients $\{A_n\}$. This system may be solved using an iterative scheme when 
$C(\mu)$ is strictly decreasing. Once solved, the particle velocity is obtained using \eqref{recip in c} and \eqref{def C}, yielding 
\begin{equation}
\mathcal{U}\approx-\frac{2}{3}MA_1\Pen^{-1/3}. \label{final U}
\end{equation}
For a given distribution $k(\mu)$ one needs in general to solve \eqref{C final} separately for $M=\pm1$, resulting in two different distributions --- say $C_\pm(\mu)$. In the case where $k(\mu)$ is an odd function, however,
it is readily verified from \eqref{C final} that $C_-(\mu)=-C_+(-\mu)$. 
Given \eqref{recip in c}, the resulting particle  velocities are opposite in sign. This, of course, is obvious by symmetry.

The asymptotic prediction \eqref{final U} can be compared with a direct computational solution of equations \eqref{advection--diffusion}--\eqref{slip}, obtained using a spectral decomposition in the azimuthal direction and a stretched radial grid (see Ref.~\cite{Michelin:14} for more details).
As an example we consider the simple distribution $k(\mu)=\mu$. Since this function is odd, it is sufficient to solve 
\eqref{C final} for $M=1$: we then get from \eqref{recip in c} that $A_1= -0.86$ whereby \eqref{final U} gives $\mathcal{U} \approx 0.57\Pen^{-1/3}$. This result is in excellent agreement with the full numerical solution: see figure~\ref{fig1}(b). The solute-concentration maps, shown in the insets,  illustrate the transition from an essentially fore--aft symmetric distribution at weak convection to a boundary-layer structure at strong convection.

Note that the use of the macroscale model of Ref.~\cite{Michelin:14} within a boundary layer of dimensional thickness $a\delta$ implies that
the underlying limit of that model, $\lambda \ll a$, should be refined to
$\lambda\ll a\delta$. Moreover, the very derivation of the effective conditions \eqref{flux} and \eqref{slip} 
actually requires $(\lambda/a) \Pen\ll1$ and $(\lambda/a)^2 \Pen\ll1$, respectively \citep{Michelin:14}. With the boundary-layer thickness $\delta$ being $\Pen^{-1/3}$ it follows that 
the stringent condition which must be satisfied is $\lambda/a \ll \Pen^{-1}$. Ref.~\cite{Michelin:14} estimated that
$\lambda/a$ is between $10^{-5}$ and $10^{-3}$, implying that the computations performed for the largest P\'eclet number employed herein ($\Pen=100$) still fall within the validity domain of the model.

\section{Fixed-rate model}
We now consider the fixed-rate model, where the  rate-constant  $\mathscr{K}$ of a first-order
kinetic relation is specified on the particle boundary (and assumed axially symmetric).  
Since a characteristic rate of solute absorption is $\|\mathscr{K}\|\mathscr{C}_\infty$, where $\| \mathscr{K} \|$ is a characteristic norm of $\mathscr{K}$, 
the velocity scale \eqref{u scale} is replaced with 
\begin{equation}
\frac{\| \mathscr{K} \| \mathscr{C}_\infty kT \lambda^2}{\eta D}. \label{u scale rate}
\end{equation}
The P\'eclet number is now of order (cf.~\eqref{Pe in A}) 
\begin{equation}
\frac{a\| \mathscr{K} \| \mathscr{C}_\infty  kT \lambda^2}{\eta D^2}. \label{Pe scale}
\end{equation}

The limit of short-range interaction is again governed by \eqref{advection--diffusion}--\eqref{slip}, except that \eqref{flux} is replaced by
\begin{equation}
\pd{c}{r} = k(\mu)(1 + \Da \, c) \quad \text{at}\quad  r=1 \label{flux rate}
\end{equation}
wherein
\begin{equation}
\Da = \frac{a\| \mathscr{K} \|}{D}. \label{Da}
\end{equation} 
is the Damk\"ohler  number and now $k=\mathscr{K} / \| \mathscr{K} \|$, which is again an $O(1)$ distribution; since the first-order kinetic describes solute adsorbing onto the boundary, $k$ is here non-negative.

When addressing here the limit of large P\'eclet numbers care must be exercised. Given \eqref{Pe scale}, one would typically envision large $\Pen$ due to fast reaction (large $\| \mathscr{K} \|$) or large swimmers (large $a$). Since however  both $\Pen$ and $\Da$ are linear in $a\| \mathscr{K} \|$, the proper limit to consider is that where $\Da$ becomes $O(\Pen)$.

With a large P\'eclet number, one may na\'ively expect the topology of the fixed-flux analysis, namely the trivial solution \eqref{c is 0} with a non-zero excess concentration confined to a thin boundary layer. 
A dominant-balance inspection reveals however that no boundary-layer structure is compatible with  condition \eqref{flux rate}. This apparent paradox is readily resolved by noting that, if 
\begin{equation}
\bu=O(\Pen^{-1})  \text{ or asymptotically smaller}, \label{u small} 
\end{equation}
the advective term does \emph{not} dominate
\eqref{advection--diffusion}. Approximation \eqref{advection dominated} is then rendered invalid, as is then \eqref{c is 0}; thus, no boundary layer is realized.  

We therefore  proceed under the  \textit{a priori} assumption \eqref{u small}. The scaling of $c$ and $\bu$ are accordingly obtained by considering the dominant balances of \eqref{flux rate} in the absence of a boundary layer. As it turns out, the only consistent balance is that  between the last two terms in \eqref{flux rate}, implying that $c$ is $O(\Da^{-1})$. Note however that this is not the velocity scaling, since the present balance implies a uniform leading-order value of $c$ (namely $-1/ \Da$) on the boundary, so the slip is triggered by the leading-order correction to the $O(\Da^{-1})$ concentration. It follows from \eqref{flux rate} that this correction is of order $\Da^{-2}$, thus providing the velocity scaling. With $\Da=O(\Pen)$, this is indeed compatible with  \eqref{u small}. 

With the fluid velocity being $O(\Pen^{-2})$ it becomes evident that the leading-order transport is (counter-intuitively)  unaffected by advection. Following the preceding arguments we postulate the expansions 
\begin{equation}
c = \Da^{-1} c_1 + \Da^{-2} c_2 + \cdots
, \quad
\bu= \Da^{-2} \bu_2 + \cdots, \quad \mathcal{U}= \Da^{-2}\mathcal{U}_2 + \cdots. 
\end{equation}
At $O(1)$, condition \eqref{flux rate} yields 
\begin{equation}
c_1 = -1 \quad \text{at} \quad r=1 .\label{c1 r=1}
\end{equation}
From \eqref{advection--diffusion} we find that the leading-order concentration $c_1$ is harmonic. The solution that satisfies \eqref{c1 r=1} and decays at infinity is the monopole
\begin{equation}
c_1= -\frac{1}{ r}. \label{mono}
\end{equation}
At $O(\Da^{-1})$, condition \eqref{flux rate}   reads
\begin{equation}
\pd{c_1}{r} =  k(\mu) c_2 \quad \text{at} \quad r=1 .\label{c2 r=1}
\end{equation}
Substitution of \eqref{mono} yields $c_2 = 1/k(\mu)$ at $r=1$. Note that there is no need to solve for $c_2$ in the fluid domain, where it is governed  by an advection--diffusion equation. 
Indeed, using \eqref{recip in c} we readily obtain
\begin{equation}
\mathcal{U}_2 = -M \int_{-1}^1\frac{\mu\, \dd \mu}{k(\mu)}. \label{asymp U}
\end{equation}

\begin{figure}
\centering
\includegraphics[scale=0.45]{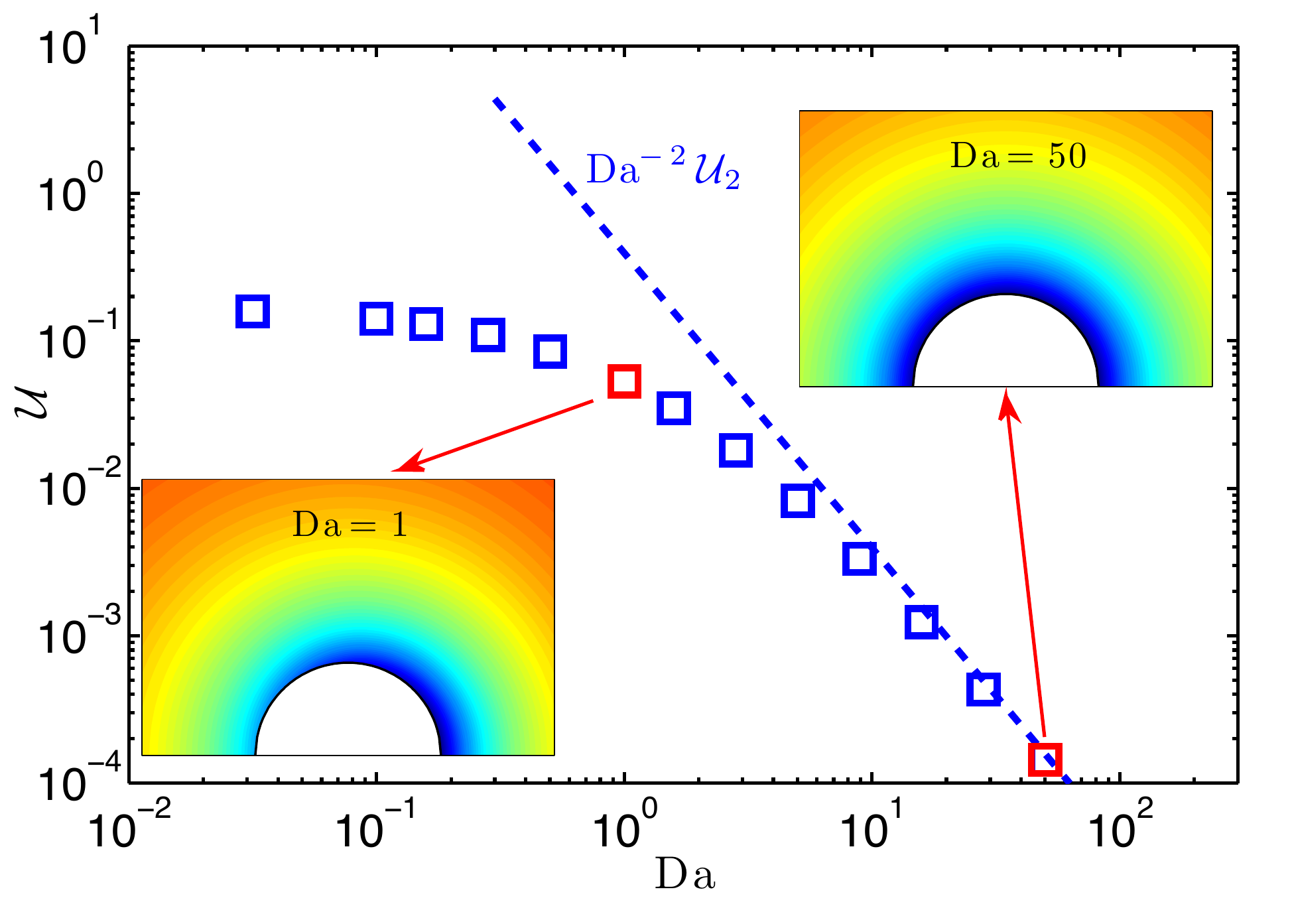} 
\caption{Particle velocity $\mathcal{U}$ for the fixed-rate model, with $k(\mu)=1+\mu/2$ and $M=1$. Line: large-$\Pen$ approximation; symbols: computational results, performed for $\Da=\Pen$. The insets depict the solute-concentration distribution (scaled by the maximum surface concentration) as provided by the numerical solutions for $\Da=1$ and $\Da=50$. Note the approach to a spherically symmetric distribution with increasing $\Pen$ (cf.~\eqref{mono}).}
\label{fig2}
\end{figure}

In figure \ref{fig2} we compare the  asymptotic approximation $\mathcal{U} \approx \Da^{-2}  \mathcal{U}_2$ with the computational solution. This is done for $M=1$ and the rate-constant distribution $k(\mu)=1+\mu/2$, for which \eqref{asymp U} yields $\mathcal{U}_2 = 2\ln9-4 \approx 0.3944$. The computations were performed for $\Da=\Pen$. At large $\Da$ they indeed agree with the asymptotic approximation. The solute-concentration maps, shown in the insets, illustrate the transition at large $\Da$ to the radially symmetric distribution \eqref{mono}
\section{Concluding remarks}
We have analysed self propulsion of a chemically reactive particle for large P\'eclet numbers. 
In the fixed-flux model, the excess-solute concentration is localised in a narrow boundary layer. Use of boundary-layer approximations reduces the coupled problem governing the nonlinear solute transport and fluid motion to the solution of an integral equation governing the interfacial solute concentration. In the fixed-rate model, we focused upon the limit of large swimmers or strong reaction, where the Damk\"ohler number becomes comparable to the P\'eclet number. In that problem no boundary layer occurs, and solute advection actually diminishes with increasing $\Pen$.

The respective scaling of the dimensionless velocity with $\Pen^{-1/3}$ and $\Da^{-2}$ implies that the velocities \eqref{u scale} and \eqref{u scale rate}, characteristic of the flow at moderate $\Pen$, are no longer representative at large $\Pen$. Rather, making use of 
\eqref{u scale} and \eqref{Pe in A} in the fixed-flux case and \eqref{u scale rate} and \eqref{Da} in the fixed-rate case, we find instead the respective velocity scales (cf. Ref.~\cite[][]{Julicher:09:generic})
\begin{equation}
\frac{\| \mathscr{A} \|^{2/3} (kT)^{2/3} \lambda^{4/3}}{\eta^{2/3}D^{1/3}a^{1/3}}, \quad
\frac{ \mathscr{C}_\infty kT \lambda^2 D}{\eta a^2 \|  \mathscr{K} \|},
\label{new u scale}
\end{equation}
representing a transition from a size-independent velocity to ones that decrease with particle size. With the generic velocity scales  \eqref{u scale} and \eqref{u scale rate} being non-representative at large $\Pen$, it follows that the dimensionless number $\Pen$ itself, as provided by \eqref{Pe in A} and \eqref{Pe scale}, does not  constitute a  
{genuine} P\'eclet number,  but simply a measure of the surface activity. This is  reminiscent of electrokinetic phenomena at strong applied fields \citep*{Schnitzer:12:strong,Schnitzer:13:drop,Schnitzer:14:bubble} 
or due to imposed flows \citep*{Yariv:11}, where the generic P\'eclet  number defined by Ref.~\cite{Saville:77} no longer  represents the relative magnitudes of advection and diffusion. 


The present work suggests two future directions. The first is a different analysis of the fixed-reaction-rate model, appropriate to the case where $\Pen$ becomes large due to large solute molecules (small $D$). In that scenario, where $\Da$ is $O(\Pen^{1/2})$, the boundary-layer thickness is $\Pen^{-1/2}$ with an identical scaling for the velocity field (so \eqref{u small} is satisfied). The boundary-layer problem is then similar to that formulated in \S\/\ref{subsec:Boundary-layer formulation}, except that \eqref{flux C} is replaced by an inhomogeneous Robin condition.

The other direction involves the case where the chemical reactions produce ions, rather than neutral species. In that case the interaction layer is the Debye diffuse-charge layer. This problem is fundamentally different, as the interaction potential (namely the electric potential) is itself coupled to the ionic concentrations through Poisson's equation. This ``auto-electrophoresis'' problem therefore falls into the realm of electrokinetics \citep*{Moran:10}. The macroscale description of that problem, appropriate to the limit of thin double layers, was developed by Ref.~\cite{Yariv:11:self-propulsion}. It 
differs from the macroscale description of Ref.~\cite{Michelin:14} in several fundamental aspects, the important one being 
the need to solve for the nonlinearly-coupled transport of two fields (ionic concentration and electric potential).
Solute advection then constitutes only one of several nonlinearities inherent in the problem. The macroscale model of Ref.~\cite{Yariv:11:self-propulsion} was solved in that paper using a linearization scheme, appropriate to the case of a nearly homogeneous particle. It is desirable to extend this solution with numerical computations, similar to those of Ref.~\cite{Michelin:14}, as well as a large-P\'eclet-number asymptotic analysis, comparable to that appearing in the present paper. 
\begin{acknowledgments}
EY was supported by the Israel Science Foundation (grant no.~184/12). SM acknowledges the support of the French Ministry of Defence through a DGA grant.
\end{acknowledgments}

\bibliographystyle{jfm} 

\end{document}